\documentclass[amsfonts, prd, 
nofootinbib, showpacs]{revtex4}
\usepackage[utf8]{inputenc}
\usepackage{amssymb}
\usepackage{amsmath}
\usepackage{amsfonts}
\usepackage{graphicx, float}
\usepackage{graphicx, epsfig}
\usepackage{color}
\usepackage{enumerate}
\newcommand{\beq}{\begin{equation}}
\newcommand{\eeq}{\end{equation}}
\newcommand{\bea}{\begin{eqnarray}}
\newcommand{\eea}{\end{eqnarray}}

\newcommand{\Mpl}{M_{\rm Pl}}
\newcommand{\Madm}{M_{\rm ADM}}
\newcommand{\ellp}{\ell_{\rm Pl}}

\newcommand{\RC}{r_{\rm C}}
\newcommand{\RS}{r_{\rm S}}
\newcommand{\Tpl}{T_{\rm Pl}}

\newcommand{\Ma}{M_\mathrm{ADM}}

\begin{document}
\title{Self-complete and GUP-Modified Charged and Spinning Black Holes}

\author{Bernard Carr$^1$\footnote{E-mail: b.j.carr@qmul.ac.uk}, Heather Mentzer$^{2,3}$\footnote{E-mail: hmentzer@ucsc.edu}, 
Jonas Mureika$^2$\footnote{E-mail: jmureika@lmu.edu}, Piero Nicolini$^{4,5}\footnote{E-mail: nicolini@fias.uni-frankfurt.de}$}
\affiliation{$^1$School of Physics and Astronomy, Queen Mary University of London, Mile End Road, London E1 4NS, UK\\
$^2$Department of Physics, Loyola Marymount University, Los Angeles, CA\\
$^3$Department of Physics, University of California Santa Cruz, Santa Cruz CA\\
$^4$Frankfurt Institute for Advandced Studies (FIAS), Frankfurt am Main, Germany\\
$^5$Institut f\"ur Theoretische Physik, Johann Wolfgang 
Goethe-Universit\"at, Frankfurt am Main, Germany}


\begin{abstract}
We explore some implications of our
previous proposal, motivated in part by the Generalised Uncertainty Principle (GUP) 
and the possibility that black holes have  quantum mechanical hair 
that  the ADM mass of a system has the form $M + \beta \Mpl^2/(2M)$, where $M$ is the bare mass, $\Mpl$ is the Planck mass and $\beta$ is a positive constant. This also suggests some connection between black holes and elementary particles and supports the suggestion that gravity is self-complete. 
We extend our model to charged and rotating black holes, since this is clearly relevant to elementary particles. 
The standard Reissner-Nordstr\"om and Kerr solutions 
include zero-temperature states,
representing the smallest possible black holes, and already exhibit features of the GUP-modified Schwarzschild solution.   
However, interesting new features arise if the charged and rotating solutions are themselves GUP-modified. In particular, there is an interesting transition below some value of $\beta$ from the GUP solutions (spanning both super-Planckian and sub-Planckian regimes) to separated super-Planckian and sub-Planckian solutions. 
Equivalently, for a given value of $\beta$, there is a critical value of the charge and spin above which the solutions bifurcate into sub-Planckian and super-Planckian phases, separated by a mass gap in which no black holes can form.
\end{abstract}



\maketitle

\section{Introduction}

\begin{figure}[h!]
\includegraphics[scale=.5]
{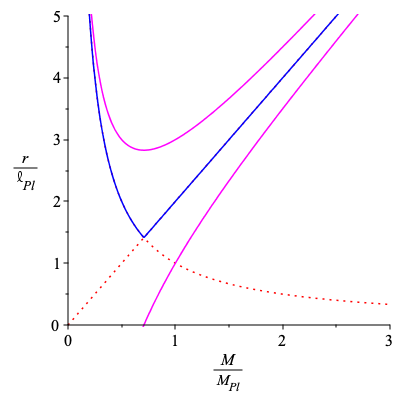}
\caption{
The blue lines show 
the Schwarzschild and Compton scales, the Schwarzschild solution being self-complete in the sense that the intersection gives 
the smallest resolvable length scale.
The red dotted lines give the 
(inaccessible) continuations of these
 curves. The upper 
magenta curve shows the unified
Compton and Schwarzschild scale ($r_{\rm CS}$) 
 if the BHUP correspondence with $\beta > 0$ applies. The lower 
magenta curve applies for $\beta < 0$ but this does not provide a unification.}
 \label{SCfig}       
\end{figure}

Any final theory of physics must amalgamate quantum theory, which applies in the microscopic domain, with general relativity, which applies in the macroscopic domain. 
Key features of these regimes are the (reduced) Compton wavelength, $\RC = \hbar/(Mc)$, relevant to particles, and the Schwarzschild radius, $\RS = 2GM/c^2$, relevant to black holes. 
As shown by the blue curves in Figure~\ref{SCfig}, these two length scales intersect where
\beq
r_{\rm S} = r_{\rm C} ~~~\Longrightarrow~~~
 M_{\rm min} = \Mpl/\sqrt{2} , ~~~
r_{\rm min} = \sqrt{2} \, \ellp \, ,
\label{SC}
\eeq
where $\ellp = \sqrt{\hbar G/c^3} \sim 10^{-33}$cm and $M_{\rm Pl} =\sqrt{\hbar c/G} \sim 10^{-5}$g  are the Planck scales at which quantum gravity becomes significant. This has the important implication that 
any attempt to probe a particle above the Planck energy will result in the formation of a black hole, so that one probes the Schwarzschild radius instead. This is referred to as the `self-completeness' of gravity \cite{dvali10a,dvali10b,dvali11,Adler_4,piero4,mureika13,aurilia13a,aurilia13b},
although the precise 
meaning of this term will be modified as a result of the considerations of this paper. 

Of course, one would not expect the standard expressions for $\RS$ and $\RC$ to apply all the way down to the Planck scale, so Eq.~(\ref{SC}) is questionable. For example, as one approaches the Planck point from the left, it has been argued~\cite{Adler_1, Adler_2, Adler_3}
that the Heisenberg Uncertainty Principle (HUP) should be replaced by a Generalized Uncertainty Principle (GUP), which corresponds to  a generalized reduced Compton wavelength of the form 
\beq
\RC' 
= \frac{\hbar}{Mc} \left[ 1  + \alpha  \left( \frac{M}{\Mpl} \right)^2 \right] ~~~ (M < \Mpl) \, ,
\label{RC}
\eeq 
where $\alpha$ is a dimensionless constant. On the other hand, as one approaches the intersect point from the right, it has been argued that  the Schwarzschild expression should be replaced by a generalized event horizon (GEH) of the form
\beq
 \RS' = \left( \frac{2 GM}{c^2} \right) \left[ 1  + \frac{\beta}{2} \left( \frac{\Mpl}{M} \right)^2 \right] ~~~ (M > \Mpl)\, ,
 \label{RS}
\eeq
 where $\beta$ is another dimensionless constant. For example, this is expected in the N-portrait model of  
Dvali {\it et al.} \cite{dvali}.
The condition $\RS' = \RC'$ then gives 
\beq
M_{\rm min} = 
 \sqrt{\frac{\beta-1}{\alpha - 2}} \, M_{\rm Pl} \,, \quad r_{\rm min} = \frac{2 - \alpha  \beta}{\sqrt{(\alpha - 2)(\beta -1)}} \, \ellp \, , 
\label{RS'}
\eeq
which reduces to Eq.~(\ref{SC}) for $\alpha =  \beta = 0$. 
However, Eqs.~(\ref{RC}) and (\ref{RS}) might 
merely give the lowest order terms in a more precise theory, in which case Eq.~\eqref{RS'} would be modified.

Although the self-completeness condition circumvents the pathology of the singularity in the Schwarzschild metric, the discontinuity at $M_{\rm min}$ corresponds to a critical point which represents some type of phase transition between black holes and particles.  The implication is that $M_{\rm min}$ corresponds  to both the lightest possible black hole and the heaviest possible particle. 
However, the similarity of Eqs.~(\ref{RC}) and (\ref{RS}) suggests another  view, in which there is some deep connection between the Uncertainty Principle  (which underlies the Compton expression) on small scales and black holes on large scales, so that there is a smooth transition between the  two expressions. This is termed the Black Hole Uncertainty Principle (BHUP) or Compton-Schwarzschild (CS) correspondence \cite{CMP,carr2,lake} and is manifested in a unified expression for the Compton wavelength
on sub-Planckian mass scales ($M < \Mpl$) and the Schwarzschild radius
 on super-Planckian (sometimes termed trans-Planckian) mass scales  ($M > \Mpl$). 
So the issue is whether there is some way of merging the 
expressions for $\RC' $ and $\RS'$. 

The simplest way to amalgamate Eqs.~\eqref{RC} and \eqref{RS} would be to put $\alpha =2$ and $\beta =1$,  so that  the curves are identical rather than intersecting. However, a model with no free parameters has little appeal to either GUP or GEH advocates. A more
natural (one-parameter) amalgamation is 
\beq
r_{\rm CS} = \frac{\beta \hbar}{Mc} + \frac{2GM}{c^2} \, ,
\label{unified}
\eeq
as illustrated by the upper curve in Fig.~\ref{SCfig}. This has a smooth minimum for $\beta > 0$ and is formally equivalent to Eq.~(\ref{RS}), except that it applies  for both $M < \Mpl$ and $M > \Mpl$.  
Eqn~\eqref{unified} is not formally equivalent to Eq.~(\ref{RC}) because the free parameter is associated with the first term rather than the second. However, this seems plausible  
because the Compton scale arises in various physical contexts, each corresponding to a different value of $\beta$ \cite{lake}.  It would be unnatural to associate the free parameter with the second term because the expression for the Schwarzschild radius is exact. Note that the coefficient in the Heisenberg Uncertainty Principle is precise, since $\Delta x = \hbar/(2 \Delta p)$, but the issue is how ones goes from $\Delta x$ and $\Delta p$ to $\RS$ and $M$. 

In our previous paper \cite{CMN}, we suggested a simple realization of this proposal, in which the ADM mass of a system  ({\it i.e.} the mass measured gravitationally at large distances) is related to the bare mass $M$ by
\beq
M_{\rm ADM} 
= M +  \frac{\beta \Mpl^2}{2M}
\label{madm}
\eeq
 for some positive constant $\beta$. Thus $M_{\rm ADM} \approx M$ for $M \gg \Mpl$ but scales as $1/M$ for $M \ll \Mpl$ and has a minimum value of 
 $ \sqrt{2 \beta} \, \Mpl$ at $M = \sqrt{\beta/2} \, \Mpl$. We described this as the `$M + 1/M$' model  and 
it might be motivated by the approach of Dvali {et al.} 
cited above, with the $1/M$ term being be regarded as quantum mechanical hair.  It may be argued that $M$ is related to the invariant energy $\sqrt{s}$ for a hypothetical collision at the Planck energy.
Note that $r_{\rm CS}$ only has a smooth minimum in  Fig.~\ref{SCfig} for $\beta > 0$.  
For $\beta < 0$,
it reaches $0$ at $M = \sqrt{|\beta|/2} \, \Mpl$, as illustrated by the lower (magenta) curve in Fig.~\ref{SCfig}, but there is no Compton curve on the left because $r_{\rm CS}$ is negative. Since $ \RS' = 0$ at this point, one effectively has $G \rightarrow 0$  (no gravity), which relates to models involving asymptotic safety \cite{bonnano}.  
For the rest of this paper we focus on the $\beta>0$ case. 

There are other possible forms for $r_{\rm CS}$, more complicated than Eq.~(\ref{unified}), which asymptote to $\RC$ for $M \ll \Mpl$ and $\RS$ for $M \gg \Mpl$. For example, whereas Eq.~(\ref{unified}) corresponds to a {\it linear} GUP, one could consider {\it quadratic} forms ($M_{\rm ADM} \sim \sqrt{M^2 + 1/M^2}$), 
such as arise in Loop Quantum Gravity \cite{CMP}.  Whatever the form, this suggests 
some connection between black holes and elementary particles, with the sub-Planckian black holes having a size  of order the Compton wavelength for their mass. In this case, the distinction between an elementary particle and a black hole, assumed in the original formulation of gravitational self-completeness, no longer applies. 
The proposal that elementary particles could be black holes originally arose in the context of the `strong' gravity model of the 1970s \cite{holzhey2,holzhey3} and was  motivated by the similarity of the  $J(M)$ relations for hadrons (viz. their Regge trajectories) and extreme rotating black holes \cite{holzhey1}. Of course, with standard gravity, an elementary particle of mass $m$ is larger than its Schwarzschild radius by a factor of $(\Mpl/m)^2$, this being $10^{38}$ for a proton, so it could only be a black hole if the strength of gravity were increased by this factor.
Recently this idea has been revived in a more modern context \cite{oldershaw} and it might also be associated with the effects of extra compact dimensions \cite{carr1}.

In this paper we extend our previous analysis to the Reissner-Nordstr\"om (RN) and Kerr solutions.
Since most elementary particles have spin and charge, and since quantum black holes
created in ultra-high-energy environments and particle collisions are also likely to be initially charged and rotating, such an extension is very natural and indeed required \cite{burinskii}. 
The charge will be an integer multiple of the electron charge 
and the spin will be a multiple of Planck's constant.  Determining how much charge and spin each
solution allows 
is conducive to a better understanding of the nature of these quantum black holes.  
Although the first part of our analysis does not explicitly invoke the BHUP correspondence,
we will show that the charged and rotating solutions exhibit features of  
the `$M+1/M$' solution if they are far from extremal. 

We then turn to the $`M+1/M'$ solutions themselves and address two aspects of the Planck-scale behaviour: self-completeness and possible BHUP modifications to the metric.
 In the first case, we are interested in determining the
minimum (maximum) possible  mass of a black hole (particle) such that 
solutions are self-complete.   
In the second case, we modify the RN and Kerr metrics subject to the $M \rightarrow M + 1/M$ 
correction and calculate the associated thermodynamic quantities.  For RN black holes, we find that the BHUP modification allows for complete evaporation ($T=0$) in the $M=0$ limit, so long as the charge of the black hole is small.  In the extremal limit, the temperature profile bifurcates to admit a classical sub-Planckian black hole, as well as a new sub-Planckian object.  Although these solutions are of great physical interest, it should be stressed that they exhibit the sort of mass gap which the BHUP correspondence was originally intended to remove. 

The plan of this paper is as follows. Sec.~II reviews our previous work, describing self-complete and GUP-modified Schwarzschild black holes. Sec.~II discusses the standard self-complete  RN black hole. The $M+1/M$ (GUP-modified) version of this is then discussed in Sec.~III and
reveals similar behaviour to the Schwarzschild $M+1/M$ case, at least for a suitable range of parameters. Sec.~IV discusses the standard self-complete and GUP-modified Kerr solutions, the results being qualitatively similar to the charged case. We draw some general conclusions in Sec.~V, with particular emphasis on self-completeness and the link beween black holes and elementary particles. 
 
\section{Self-Complete and GUP-Modified Schwarzschild Black Holes}

The simplest way of implementing the BHUP correspondence is to use the 
GUP-modified Schwarzschild metric obtained in Ref.~\cite{CMN}:
 \bea
ds^2  = f(r) dt^2 - f(r)^{-1} dr^2 - r^2 d\Omega^2
\label{RN}
\eea
with
\bea
f(r) = 1-\frac{2M_{\rm ADM}}{\Mpl^2r} \,,  \quad d\Omega^2 = d \theta^2 + \sin^2 \theta \,  d \phi^2 \, ,
\label{newmetric}
\eea
where $M_{\rm ADM}$ is given by Eq.~(\ref{madm}) and henceforth we use units with $G = M_{\rm Pl}^{-2}$ and $\hbar = c =1$ throughout this paper.  This modification to the metric ensures that the event horizon radius is given by Eq.~(\ref{unified}), allowing the possibility of both the standard super-Planckian black holes with $ r_{\rm CS} = 2M/\Mpl^2$
 and  sub-Planckian black holes with $r_{\rm CS} = \beta/M$. 
  It also smooths out
the self-complete discontinuity. 
There is an interesting connection here
with the quantum N-portrait model of Dvali {\it et al.} \cite{dvali1,dvali2,dvali3,dvali4,dvali5}, which regards a black hole as a weakly-coupled Bose-Einstein condensate of gravitons.  From holographic considerations, the number of gravitons (entropy states) in the black hole is
\beq
N \approx \frac{A_{\rm BH}}{\ellp^2}
  \approx \frac{M^2}{\Mpl^2} \, ,
\eeq
where $A_{\rm BH}$ is the black hole area. As noted
in Ref.~\cite{frassino}, one can then argue that the black hole radius is
\beq
r_{\rm CS} \approx 
\frac{2M}{\Mpl^2} \left(1 +\frac{\beta}{2N}\right) =  \frac{2M_{\rm ADM}}{\Mpl^2} \quad (M > \Mpl) \, ,
\label{case2}
\eeq
which is equivalent to Eq.~(\ref{unified}).

\begin{center}
\begin{figure}[h!]
\includegraphics[scale=0.4]{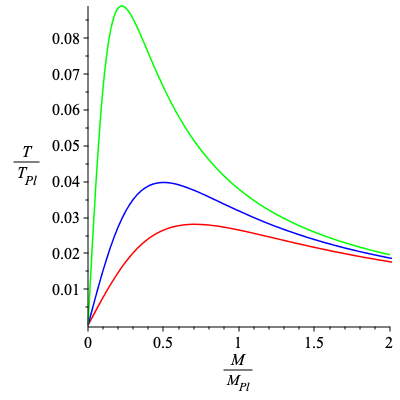}
\includegraphics[scale=.5]{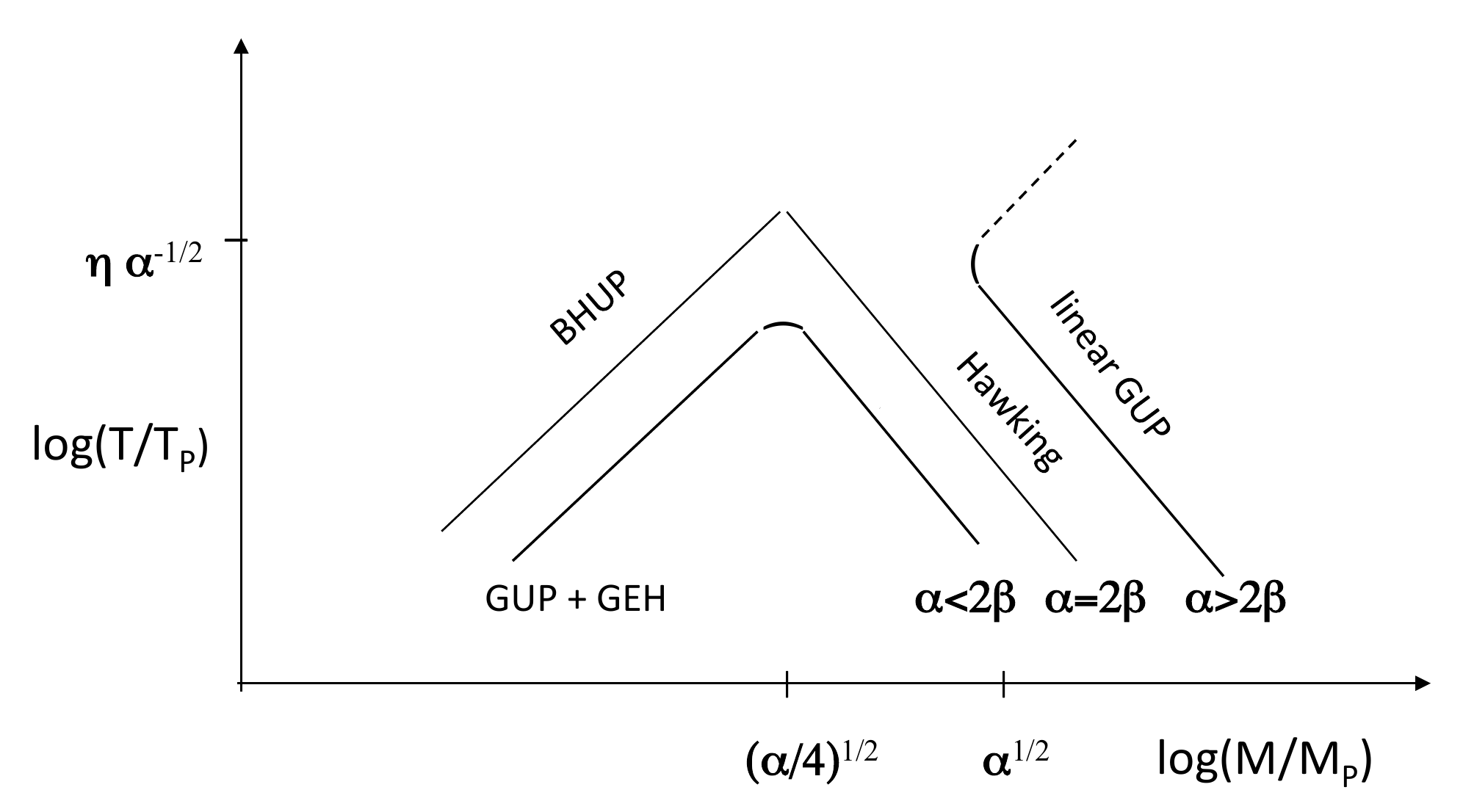}
\caption
{(a) Temperature implied by surface gravity argument as a function of mass  in `$M+1/M$' model for $\beta = 1$ (red, bottom), $\beta = 0.5$ (blue, middle) and $\beta = 0.1$ (green, top).  The temperature reaches a maximum and then decreases, so that the black hole cools to a configuration with $M=T=0$.  
(b) 
This gives the temperature in the more general case with both $\alpha$ and $\beta$ for $\alpha < 2\beta$ (cf.  `$M+1/M$'  model), $\alpha = 2\beta$ (cf.  Hawking solution) and $\alpha > 2\beta$ (cf. Adler's solution).}
\label{fig3}
\end{figure}
\end{center}

Given the metric (\ref{RN}), one can obtain the black hole temperature from its surface gravity \cite{hawking1,hawking2}:
\beq
kT = 
\frac{\kappa}{2 \pi} =
   \left. \frac{1}{4 \pi}  \frac{dF}{dr} \right|_{r_{\rm CS}} = \frac{\Mpl^2}{8\pi M(1+\beta \Mpl^2/2M^2)} \, . 
\label{hawktemp}
\eeq
This is plotted in Fig.~\ref{fig3}(a) and the limiting behaviour in the 
asymptotic regimes is 
\beq
kT \approx
\begin{cases}
\frac{\Mpl^2}{8\pi M} \left[ 1- \frac{\beta}{2}  \left( {\Mpl \over M} \right)^{2} \right]  
& (M \gg \Mpl) \\
 \frac{M}{4\pi \beta} \left[ 1- \frac{2}{\beta}  \left( {M \over \Mpl} \right)^{2} \right] 
& (M \ll \Mpl)  \, . 
\end{cases}
\label{mcases}
\eeq
The large $M$ limit is the usual Hawking
temperature but, as the black hole evaporates, the temperature reaches a maximum at $M = \sqrt{\beta/2} \, \Mpl$ and then decreases to zero as $M \rightarrow 0$. 

Before proceeding, it is useful to compare Eq.~\eqref{hawktemp} with the temperature derived with other approaches. In the standard picture, one can 
calculate the black hole temperature from the HUP by 
identifying the Schwarzschild radius with $\Delta x$ and the black hole temperature with a multiple $\eta$ of  $\Delta p$. This gives
\begin{eqnarray}
kT = \eta \Delta p = \frac{ \eta}{\Delta x} = \frac{\eta \Mpl^2}{ 2M} \, ,
\label{temper}
\end{eqnarray}
which is precisely the Hawking temperature 
if we take $\eta = 1/(4\pi)$. 
This approach can also be used to derive the black hole temperature for a model in which one adopts GUP but assumes that the expression for the black hole size is unchanged ({\it i.e.} $\beta =0$). 
In particular, Adler {\it et al.} \cite{Adler_1, Adler_2, Adler_3}
 calculate the modification required  if $\Delta p$ and $\Delta x$ are related by the linear GUP, 
\beq
\Delta x = \frac{1}{\Delta p} + \alpha \frac{\Delta p}{\Mpl^2} \, .
\eeq
Since $\Delta x$ is still
identified with the Schwarzschild radius, one obtains
\begin{equation}
kT = {\eta M  \over \alpha} \left(1\pm \sqrt{1- \frac{\alpha \Mpl^2}{M^2}} \right) \, .
\label{adlertemp2}
\end{equation}
The negative sign just gives a small perturbation to the standard Hawking temperature in the super-Planckian regime: 
\begin{equation}
kT \approx
{\eta \Mpl^2 \over 2 M}  \left[ 1 - {\alpha \Mpl^2 \over 4M^2} \right] \quad (M \gg \Mpl) \, .
\label{adler3}
\end{equation}
However, the solution becomes complex when $M$ falls below $\sqrt{\alpha} \, \Mpl$, corresponding to a minimum mass,
and it then connects to the positive branch of Eq.~\eqref{adlertemp2}. This asymptotes to $2 \eta M/\alpha$, which is presumably unphysical since it exceeds the Planck temperature.  Note that Eq.~\eqref{RS'} with $\beta=0$ gives \beq
M_{\rm min} = 
\frac{1}{\sqrt{2 -\alpha}} \, M_{\rm Pl} \,, \quad r_{\rm min} = \frac{2 }{\sqrt{2-\alpha}} \, \ellp \, , 
\eeq
so the modified Compton and Schwarzschild radii only intersect for $\alpha < 2$.

 The Adler solution has the unpalatable feature that it introduces a thermodynamic instability. 
This problem can be cured by going beyond the GUP and also modifying the relationship between the black hole radius $\Delta x$ and $M$ (i.e. by introducing the parameter $\beta$ in the expression for the generalised event horizon). This can be achieved    
by replacing $M$ in Eq.~\eqref{adlertemp2} by $M_{\rm ADM}$ and then regarding $\alpha$ and $\beta$ as independent parameters.
However, since $M_{\rm ADM}$ has a minimum value of 
$\sqrt{2 \beta} \, \Mpl$, 
 one never reaches the limiting Adler mass of $ \sqrt{\alpha} \, \Mpl$ for  $\alpha < 2 \beta$. In this case, as in the `$M+1/M$' model, the temperature reaches a maximum and then decreases 
rather than going complex. 
The dependence of $T$ on $M$ 
in the asymptotic limits can then be approximated by
\begin{equation}
kT \approx 
\begin{cases}
{\eta \Mpl^2 \over 2 M} \left[ 1 - \left( \frac {2\beta - \alpha}{4}\right)  \left( {\Mpl \over M} \right)^{2} \right]  & (M \gg \Mpl) \\
{\eta M \over  \beta} \left[ 1 - \left( \frac{2\beta - \alpha}{ \beta^2}\right)  \left( {M \over \Mpl} \right)^{2} \right]  & (M \ll \Mpl) \, .
\end{cases}
\label{sub}
\end{equation}
As expected, this is equivalent to Eq.~\eqref{mcases} if $\alpha = 0$.
The overall behaviour of $T$ is shown by the lowest curve 
in Fig.~\ref{fig3}(b). For $\alpha > 2 \beta$, $T$ has the same qualitative form as in the Adler model. 
In the special case $\alpha = 2 \beta$, the effects of the $\alpha$ and $\beta$ terms cancel and one obtains the 
solution \cite{CMP} 
\begin{eqnarray}
kT = \mathrm{min} \left[ \frac{\eta \Mpl^2}{2M} \,  , \, \frac{2 \eta M}{\alpha} \right]  \, .
\label{mess3}
\end{eqnarray}
This is indicated by the middle curve in Fig.~\ref{fig3}(b). The first expression in Eq.~\eqref{mess3} is the {\it exact} Hawking temperature, but
one must cross over to the second expression below $M = \sqrt{\alpha/4} \, \Mpl$ to avoid the temperature going above the Planck value $\Tpl = \Mpl /k$. The second expression in Eq.~\eqref{mess3}
can be obtained by putting  $\Delta x =  \alpha/(2M)$ in Eq.~(\ref{temper}).
The temperatures given by the surface gravity and GUP arguments
agree to 1st order but not to 2nd order.

 Since there are independent arguments for the GUP and GEH expressions, $\alpha$ and $\beta$ could in principle be unrelated, with  Eq.~(\eqref{RC}) applying for $M < \Mpl$ and Eq.~(\ref{RS})  for $M > \Mpl$. 
However, we note that Eq.~\ref{mcases} 
(when applied for
 $M < \Mpl$) already implies a GUP effect.
When one combines the $\alpha$ and $\beta$ terms, as in Eq.~(\ref{sub}), one therefore superposes two GUP contributions to the temperature. These cancel for $\alpha = 2 \beta$, so that the exact Hawking formula still applies, but this is not required for the BHUP correspondence. Indeed, it
seems more natural to use the `$M+1/M$' solution, 
since this removes the thermodynamic instability of the Adler solution and unifies the Compton and Schwarzschild expressions without introducing the complication of an extra degree of freedom. We therefore use Eq.~\eqref{unified} in extrapolating to the charged and rotating case and drop the parameter $\alpha$ for the rest of this paper.

\section{Self-Completeness of Reissner-Nordstr\"om Solution}
\begin{center}
\begin{figure}[b]
\includegraphics[scale=0.5]{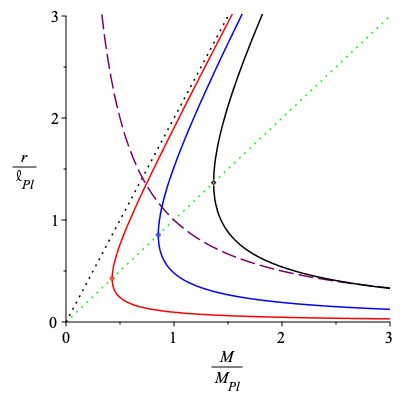}
\caption{
The solid curves show the  outer and inner horizons for a standard RN black hole with 
$n=5, 11, 16$ (left to right).  For each $n$,  the horizons meet at the extremal mass on the line $r=M/\Mpl^2$ (green dotted) and are bounded from above by the Schwarzschild radius $r_{\rm S} = 2M/\Mpl^2$ (black dotted line).  The Compton curve is shown by the dashed line and the inner horizon asymptotes to this for $n=16$.
Solutions with $n < 11$ penetrate the sub-Planckian RN regime and 
are discussed in Section~\ref{GUPRN}.}
\label{fig12}
\end{figure}
\end{center}

The RN spacetime has the well-known metric (cf. Ref.~\cite{piero18}), 
\beq
ds^2 = f(r)dt^2 - \frac{dr^2}{f(r)} - r^2d\Omega^2~~~,
\eeq
with
\beq
f(r) = 1-\frac{r_{\rm S}}{r} + \frac{r_{\rm Q}^2}{r^2}~.
\label{rnfunc}
\eeq
Here $r_{\rm S} = 2M/\Mpl^2$ and $r_{\rm Q} = Q/ \Mpl$ are characteristic gravitational and charge length scales. 
Since $Q = ne$,  where $e = \sqrt{\alpha_e}$ is the 
electron charge and $\alpha_e \approx 1/137$ is the fine structure constant, we can write the metric function as
\beq
f(r) = 1-\frac{2M}{\Mpl^2 r} + \frac{\alpha_e n^2}{\Mpl^2 r^2} \, .
\label{RN2}
\eeq
The outer (+) and inner (-) horizons are then given by
\beq
f(r_\pm) = 0~~~\Longrightarrow~~~r_\pm = \frac{M}{\Mpl^2}\left( 1\pm \sqrt{1-\frac{\alpha_e \Mpl^2 n^2}{M^2}}\right) \, .
\label{RNhorizon}
\eeq
For a black hole which is far from extremal ($ M \gg \sqrt{\alpha_e} \, n \Mpl$), this can be written as
\beq
r_\pm \approx 
\begin{cases}
\frac{2M}{\Mpl^2} \left(1 - \frac{\gamma \Mpl^2}{M^2} \right) & (+) \\ 
\frac{2\gamma}{M} \left(1 + \frac{\gamma \Mpl^2}{M^2} \right) & (-)
\end{cases}
\label{cases}
\eeq
where $\gamma \equiv \alpha_e n^2/4$.
 The form of the outer and inner 
 horizons for different values of $n$ are shown  by the upper and lower  parts of the solid curves in Fig.~\ref{fig12}, respectively. The outer horizon asymptotes to $2M/\Mpl^2$ (upper dotted curve) at large $r$ and the inner horizon to $2 \gamma /M$ at low $r$. 

For each $n$, the two horizons merge on the line  $r=M/\Mpl^2$ (lower dotted curve) at the minimum value of $M$ and have an infinite gradient ($dr/dM$) there. This corresponds to a sequence of ``extremal'' solutions (shown by the dots in Fig.~\ref{fig12}) with 
a spectrum of masses given by
\beq
1-\frac{\alpha_e \Mpl^2 n^2}{M^2}=0 ~~~\Longrightarrow ~~~ M_n = \sqrt{\alpha_e} \, n \Mpl \, .
\eeq
For given $n$, there are no solutions with  $M$ less than this since these would correspond to naked singularities. In particular, $n$ could be at most the integer part of $1/ \sqrt{\alpha_e}$ (i.e. $11 $) for a Planck-mass black hole.
It is interesting that 
Eq.~(\ref{cases}) has two asymptotic behaviors  in the $M \gg \Mpl$ regime: the outer horizon correponds to 
Eq.~(\ref{RS}) but with a negative value of $\beta$; the inner horizon corresponds to Eq.~(\ref{RC}) but with a positive value of $\alpha$
and it nearly asymptotes to the Compton wavelength for $n = 16$, this being the integer part of $\sqrt{2/\alpha_e}$.

The Compton line intersects the outer black hole horizon, as required by
the self-completeness condition  \cite{piero18}, where
\beq
r_{\rm C} = r_{+} \, 
\eeq
and this yields the mass-scale
\beq
M = \frac{\Mpl}{\sqrt{2-\alpha_e n^2}} \approx  \frac{\Mpl}{\sqrt{2- n^2/137}} \, .
\label{extreme}
\eeq
This assumes the standard definition of the Compton wavelength,
although this may be modified in the `$M+1/M$' approach. The relation between the Compton and outer horizon scales is shown in Figure~\ref{fig12}.
For $n=0$, the intersect mass is
$\Mpl /\sqrt{2}$ but it increases with $n$ and tends to $\Mpl$ as $n \rightarrow \sqrt{137}$ (middle curve). This implies a constraint $n \leq 11$ on the charge of a self-complete RN black hole. The Compton line still intersects the {\it inner} horizon for $\sqrt{137} <  n < \sqrt{274} $, with $n=16$ (right curve) being the last solution which allows this. However, these solutions  do not exhibit self-completeness since they penetrate the $r < \ellp$ region where quantum gravity applies. 
For $n > \sqrt{274 }$, not even the inner horizon intersects the Compton line, 
ensuring a clear distinction between particles and black holes.

Since most elementary particles have charge, this suggests some connection with
the BHUP correspondence, although we note that 
all known charged fundamental particles have $n = 1$. In any case,
the condition $n < \sqrt{137}$ suffices
 to account for the quantum black holes
produced in typical particle collisions ({\it e.g.} $pp \rightarrow N$, $pn \rightarrow N^\prime$ {\it etc.}).  
This creates a symmetry in the convergence to the critical point in the self-completeness diagram.  
None of these masses falls on the Compton curve and through Schwinger processes the masses make discrete jumps up towards
the $r=2M/\Mpl^2$ Schwarzschild lines.
In Section~\ref{GUPRN}, we will consider values of $n$ for curves below the Compton line, corresponding to sub-Planckian RN black holes.

\begin{center}
\begin{figure}[t]
\includegraphics[scale=0.4]{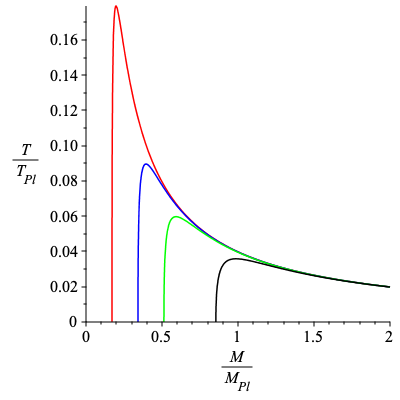}
\caption{The function $T(M)$  
for a standard RN black hole with $n=2, 4, 6, 10$ (left to right).  The curves end at $M =n \sqrt{\alpha_e} \Mpl$, corresponding to a charged remnant  with $T = 0$. }
\label{normalrn2}
\end{figure}
\end{center}

The temperature of the 
RN solution for quantized charge $Q=n\sqrt{\alpha_e}$ is calculated from the surface gravity as
\beq
kT =   \left. \frac{1}{4 \pi} \frac {df}{dr} \right|_{r_+} 
= \frac{ \Mpl^2 \sqrt{M^2 - \alpha_e n^2 \Mpl^2}}{2\pi  \, ( M + \sqrt{M^2 - \alpha_e n^2 \Mpl^2} \, )^2} \, . 
\label{rntemp}
\eeq
Figure~\ref{normalrn2} shows the function $T(M)$. It asymptotes to the Hawking expression ($ T \propto M^{-1}$) for $M \gg \Mpl$ but, as $M$ decreases, it reaches a maximum and then goes to zero as $M$ tends to the minimum mass $\sqrt{\alpha_e} \, n \Mpl$.
An important distinction between Schwarschild and RN black holes, however, is that the latter also loses charge through the Schwinger mechanism \cite{schwinger}, this operating even for extremal black holes, despite their having zero temperature. The general emission formula
 for a particle of frequency $\omega$ and charge $q$ is
\beq
\frac{dN}{dt d\omega} = \frac{\Gamma(\omega, T, q \Phi)/ 2 \pi}{\exp [(\omega + q \Phi)/T] \pm 1} \, ,
\label{Hawking}
\eeq
where $\Phi$ is the electrostatic potential and $\Gamma$ is the absorption coefficient for the relevant mode. This is equivalent to a thermal spectrum with a chemical potential proportional to the black hole charge $Q$ and  covers both
 the {\it thermal} emission associated with non-zero $T$ and the {\it athermal} emission at $T=0$.  
The thermal emission is a stochastic process, in which emitted particles can have either sign \cite{page}, whereas the athermal emission produces particles with the same sign as the black hole charge, so that the latter is always reduced.
 
Recently Lehmann {\it et al.} \cite{lehmann} have analysed this process in considerable detail  and argued that the Planck mass relics of evaporating primordial black holes are likely to be charged, thereby providing detectable dark matter candidates. 
Although we have some issues with this conclusion, primarily because Eq.~\eqref{rntemp} is not a complete representation of the Schwinger effect,
a proper analysis of this mechanism is certainly relevant for the potential identification of fundamental particles with sub-Planckian black holes. This is because
no elementary particles have charge greater than $e$, whereas 
self-complete black holes can have charge up to $11 e$. Since the Schwinger mechanism reduces the black hole charge, perhaps it can resolve this problem.

For this purpose, we recall the circumstances in which black holes are not expected to retain
charge,  as discussed by Gibbons~\cite{gibbons} and Carter~\cite{carter}. 
The electrostatic forces on a test particle with mass $m$ and charge $q$ near 
the black hole can 
overcome the gravitational pull unless
\begin{equation}
\frac{Q}{M}<\frac{m}{q} \, ,
\label{eq:Coulomb}
\end{equation}
where the masses in the present discussion are in Planck units, so that $m/q\sim 10^{-21}$ for the electron and $10^{-18}$ for the proton. This means that a black hole with just one electron
charge can retain a positron 
only if $M \gtrsim e^2/m_e \sim 10^{20}\Mpl \sim 10^{12}$ kg,
of order the mass for which the PBH lifetime is comparable to
the age of the Universe.
On the other hand, the rate per unit  volume of
electron-positron pair production through the Schwinger mechanism is
\begin{equation}
\Gamma_\mathrm{S}\simeq \frac{(e E)^2}{4\pi^3}\ e^{-E_\mathrm{c}/E} \, ,
\end{equation}
where $E_\mathrm{c}=\pi m_e^2/e$ is the critical field required for the process. Thus one requires
\begin{equation}
\frac{Q}{r_+^2}\geq \pi m_e^2/e \, .
\label{eq:Schwingertrigger}
\end{equation}
If $n=1$, this condition becomes
\begin{equation}
M \le \sqrt{\frac{\alpha_e}{4 \pi}} \, \frac{1}{m_e}  \, . 
\label{eq:Schwinger_trigger}
\end{equation}
This implies that pairs are copiously produced for
$M \lesssim e/m_e  \sim   10^{21}\Mpl \sim 10^{13}$ kg. 
Gibbons \cite{gibbons} derives a third constraint by combining Eq.~\eqref{eq:Schwinger_trigger} with the extremal condition  $Q=M$. This gives $M <  e/m_e^2 \sim 10^{43} \Mpl \sim 10^{35}$kg (i.e. $10^5 M_{\odot}$). 
In conclusion, if the black hole is not massive enough to overcome the electrostatic repulsion, it would undergo a sudden discharge not only via standard evaporation 
but also due to the Schwinger effect. 
Ruffini and colleagues have proposed that stable charged black holes could explain
gamma-ray bursts  \cite{Damour:1974qv,Preparata:1998rz} but Page has pointed out that such charged 
configurations are implausible due to the aforementioned discharge \cite{page06}. 

Our model has some similarity to the renormalization group approach of Bonanno and Reuter \cite{bonanno}, their `improved' Schwarzschild metric resembling the RN solution.  More precisely, the renormalization group equation leads to a running gravitational constant
\beq
G(k) = \frac{G_0}{1 + \omega G_0 k^2} \, ,
\eeq
where $k$ is the wave-number, $G_0$ is the Newtonian value (otherwise denoted as $G$) and $\omega$ is some constant. 
This implies a scale dependence
\beq
G(r) \approx \frac{r^3}{\gamma G_0 M} 
\eeq
 at small $r$,
leading to a $1/r^3$ correction in the potential and
\beq
f(r) \approx  1 - \frac{2G_0M}{r}\left( 1 - \frac{\omega G_0}{r^2} \right) \, .
\label{BR}
\eeq
This is similar to
 Eq.~(\ref{rnfunc}) but with a $1/r^3$ rather than $1/r^2$ term and it leads to an analogous zero-temperature extremal solution at some  critical mass  of order $\Mpl$. Hawking evaporation stops at this mass, as in the Adler model, and below it  the central singularity is either removed, leaving a smooth de Sitter core, or becomes milder. 
Such a feature is common to many
 quantum-gravity-corrected black hole models, for example, 
the non-commutative geometry inspired models \cite{nicolini06,nicolini09}, the string T-duality corrected black holes \cite{nicolini19}, the Hayward model \cite{hayward06}, the holographic screen model \cite{piero4} and a class of GUP modified metrics \cite{mijmpn,kkmn}. 

\section{GUP-Modified Reissner-Nordstr\"om Black Holes}
\label{GUPRN}

The above analysis considered  the circumstances in which the standard RN
horizon  intersects the standard Compton wavelength (the self-completeness condition). If this does not happen, there is a clear distinction between particles and black holes. Even if it does, 
Fig.~\ref{fig12} shows that there is still a discontinuity in the gradient $dr/dM$ at the intersect point, allowing the possibility of some form of  phase transition separating black holes from elementary particles. However, one does not expect either of the standard expressions to apply close to the intersect point due to quantum gravity effects. In accord with the BHUP correspondence,  we therefore  seek a smooth function $r_{\rm CM}(M)$ which asymptotes to the standard expressions for $r_{\rm C}$ for $M \ll \Mpl$ and $r_{\rm S}$ for $M \gg \Mpl$.

\begin{center}
\begin{figure}[h!]
\includegraphics[scale=0.3]{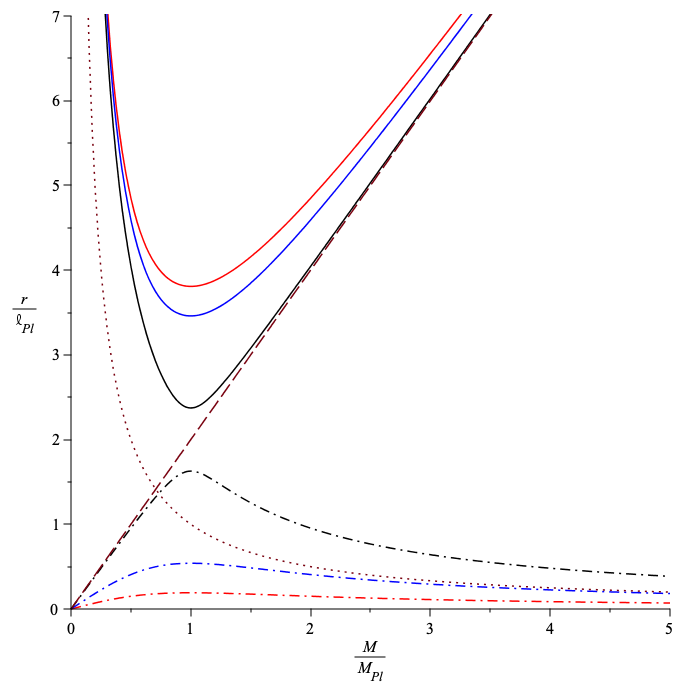}
\caption{Outer (top, solid) and inner (bottom, dash-dot) horizons for GUP-RN black holes
with $\beta = 2$ and 
$n=10$ (red), $n=16$ (blue) and $n=23$ (black).  The dashed/dotted lines show the usual Schwarzschild/Compton scales.
There is a discontinuity when $n$ exceeds 
$23$, this being the closest to the extremal solution.
 The inner horizon is nearly asymptotic to the Compton wavelength at large $M$ for $n =16$.}
\label{rnhorizon}
\end{figure}
\end{center}

As in the Schwarzschild case, we consider the  $M + 1/M$ approach, replacing $M$ by $M_{\rm ADM}= M + \beta \Mpl^2/(2M)$ in the RN metric but leaving the charge term unchanged. In principle, one  could also modify the electrostatic term 
but that would be inconsistent with the Compton wavelength of a particle not depending on its charge.
Note, however, that one no longer preserves  $M \rightarrow 1/M$ duality in the charged case with this approach, even though this was one of the original motivations for the  
Compton-Schwarzschild correspondence.
The RN
metric therefore becomes
\beq
f(r) = 1-\frac{2 M}{\Mpl^2 r}\left(1+\frac{\beta}{2} \frac{\Mpl^2}{M^2}\right) + \frac{\alpha_e n^2}{\Mpl^2 r^2} \, .
\eeq
For arbitrary $\beta$, the horizons are at
\beq
r_\pm= 
\left( \frac{M}{\Mpl^2}+\frac{\beta}{2M} \right)  \left(1\pm \sqrt{1-\frac{n^2 \alpha_e  M^2}{\Mpl^2\left(\frac{\beta}{2}+\frac{M^2}{\Mpl^2}\right)^2}}\right)  \, ,
\label{outerhor}
\eeq
which gives
the following values in the super-Planckian and sub-Planckian regimes:
\bea
M \gg \sqrt{\beta} \Mpl & ~~\Longrightarrow~~& r_+ \approx \frac{2M}{\Mpl^2} \, , ~~~ r_- \approx \frac{ \alpha_e n^2}{2M} = \frac{2 \gamma}{M} \, ,\\
M \ll  \sqrt{\beta} \Mpl & ~~\Longrightarrow~~& r_+ \approx \frac{\beta}{M} \, ,~~~ r_- \approx \frac{ \alpha_e n^2 M}{2\Mpl^2} = \frac{ 4 \gamma M}{\beta \Mpl^2} \, .
\eea
\noindent The form of $r_+$ is shown by the upper (solid) curves in Fig.~\ref{rnhorizon} for $\beta =2$. 
Since $r_+$ is a monotonic function of $M_{\rm ADM}$,
its minimum occurs at the minimum of $M_{\rm ADM}$, which corresponds to
\beq
M = M_{\rm crit} \equiv \sqrt{\beta/2} \, \Mpl  \, , \quad  r_+ = [\sqrt{2 \beta} + \sqrt{ 2 \beta - n^2 \alpha_e} \, ]  \,  \ellp \, .
\label{minhor}
\eeq
The form of $r_-$ is shown by the lower (dash-dotted) curves in Fig.~\ref{rnhorizon} for $\beta =2$, 
with a maximum at
\beq
M = M_{\rm crit} \equiv \sqrt{\beta/2} \, \Mpl  \, , \quad  r_- = [\sqrt{2 \beta} - \sqrt{ 2 \beta - n^2 \alpha_e} \, ]  \,  \ellp \, .
\eeq
Thus it occurs at the same value of $M$ as the minimum but at a smaller value of $r$.
Note that the Compton-Schwarzschild correspondence might suggest that the Compton wavelength {\it becomes} $\beta/M$  in this case, which would also modify the self-completeness condition.

For a given value of $\beta$, the square root term in Eq.~\eqref{outerhor} is real at the minimum 
or maximum for
\beq
n \leq n_{\rm max} = [\sqrt{2\beta/ \alpha_e} \, ] \, ,
\label{mext2}
\eeq
where square brackets denote the integer part. As $n$ increases, the minimum of $r_+$ decreases and the maximum $r_-$ increases until they meet 
 when $n$ reaches $n_{\rm max}$ (corresponding to the extremal case). 
Equivalently, for a given value of $n$, there is a minimum value of $\beta$ and a minimum (sub-Planckian) value of $M_{\rm crit}$:
\beq
\beta \geq \beta_{\rm min} \equiv \frac{1}{2} \alpha_e n^2 \, \Rightarrow M_{\rm crit} \geq  \frac{n\sqrt{\alpha_e}}{2}\Mpl \approx 0.043 \, n \Mpl \, .
\label{mext3}
\eeq
Table~\ref{tab1} shows the
value of $\beta_{\rm min}$ and $M_{\rm crit}$ for different values of $n$. The value of  $n_{\rm max}$ for each value of $\beta$ in the middle column can also be inferred from the entries in the left column.  Conditions \eqref{mext2} and \eqref{mext3} are required if we wish to extend the BHUP correspondence to charged black holes.
\begin{table}[h]
\begin{tabular}{lcc}
$n~~~$ & $\beta_{\rm min}$ &$M_{\rm crit}/\Mpl $  \\ \hline
2 &0.015 & 0.086 \\
10& 0.365& 0.430\\
12 &0.526& 0.516\\
14& 0.715&0.602\\
16&0.934& 0.688\\
18& 1.182& 0.774\\
20&1.460& 0.860\\
 22& 1.767& 0.946\\
 24& 2.102& 1.032\\ \hline
\end{tabular}
\caption{This shows the minimum value of $\beta$ required for $r_+$ to have a smooth minimum  for a given value of the charge $n$.  If this condition is violated, the solution has the RN form with the black hole having the minimum mass indicated.  }
\label{tab1}
\end{table}

\begin{center}
\begin{figure}[h!]
\includegraphics[scale=0.25]{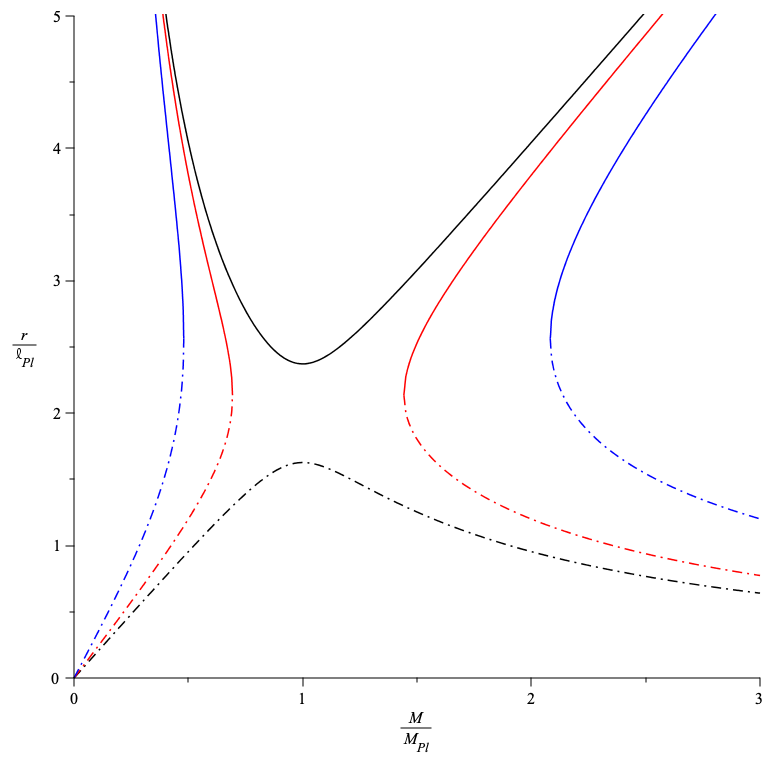}
\caption{Outer (solid) and inner (dash-dot) horizon size for GUP-RN black hole with $\beta =2$, showing the pre- ($n=23$, black) and post- ($n=25$, blue, and $n=30$, red) phase transition behaviour. The horizons for $n>23$ reach  their maximum (left) or minimum (right)
size at $M_1$ and $M_2$, respectively. There are no black holes in the mass-gap between these values.}
\label{PT_RNH}
\end{figure}
\end{center}

For $\beta < \beta_{\rm min}$ for given $n$, or $n > n_{\rm max}$ for given $\beta$, the solution has the form indicated in Fig.~\ref{PT_RNH}. There are now two branches, a RN-type solution above $\Mpl$ and a particle-type solution below $\Mpl$, with a mass gap in between. The significance of the sub-Planckian branch is unclear but it can be obtained from the super-Planckian branch by replacing $M$ with the dual mass $ \gamma \Mpl^2/M$.  Although these solutions do not exhibit continuity between particles and black holes, they are clearly of physical interest and we  examine them in more detail below. This shows that $\beta = \beta_{\rm min}$ marks a transition from the BHUP form, 
naturally linking particles and black holes, to the RN form, with a clear distinction between them.  Thus the expression for $M_{\rm crit}$ in
Eq.~(\ref{mext3}) specifies the {\it minimum} possible mass for the RN solution for given $n$. 

The novel behaviour in Fig.~\ref{PT_RNH}
can be explained as follows.  From Eq.~\eqref{outerhor}, horizons exist only for 
\begin{equation}
\Ma\geq \sqrt{\alpha_e}n\Mpl \quad \Rightarrow \quad \frac{M^2}{\Mpl}-\sqrt{\alpha_e}n M+\frac{\beta}{2} \Mpl\geq 0 \, .
\end{equation}
The sign of the discriminant, 
\begin{equation}
\xi \equiv \alpha_e n^2 -2\beta,
\end{equation}
gives  three possible situations.
For $\xi < 0$,  
there is an horizon for all values of $M$. This means that sub-Planckian black holes are admissible but none of them is extremal.
To find the minimal size of the event horizon, one can use the chain rule:
\begin{equation}
0=\frac{dr_+}{dM}=\left(\frac{dr_+}{d\Ma}\right)\left(\frac{d\Ma}{dM}\right) \, .
\end{equation}
The factor $dr_+/dM_{\rm ADM}$ is non-zero for $\Ma> \sqrt{\alpha_e}n \, \Mpl$  and
\begin{equation}
\frac{d\Ma}{dM}=1-\frac{\beta}{2}\frac{\Mpl^2}{M^2} \, ,
\end{equation}
so there is a minimal (non-extremal) horizon radius for the mass $M_{\rm crit}$ indicated by Eq.~\eqref{minhor}.
For $\xi > 0$, 
there are horizons for
\begin{eqnarray}
M< M_1&=&\frac{\Mpl}{2}\left(\sqrt{\alpha_e}n-\sqrt{\alpha_en^2-2\beta}\right)\nonumber\\
M> M_2&=&\frac{\Mpl}{2}\left(\sqrt{\alpha_e}n+\sqrt{\alpha_en^2-2\beta }\right) \, ,
\label{m1m2}
\end{eqnarray}
with $M_1<M_\mathrm{crit}<M_2$.
Thus there is a mass gap $M_1<M<M_2$ with no black holes, as observed in Fig.~\ref{PT_RNH}.
The values $M_1$ and $M_2$ correspond to extremal solutions. 
For $\beta\ll \frac{1}{2}\alpha_en^2$ one has
\begin{eqnarray}
M_1 \approx \frac{\Mpl}{2}\frac{\beta}{\sqrt{\alpha_e}n} \,, \quad 
M_2 \approx&\sqrt{\alpha_e}n \Mpl \, .
\end{eqnarray}
For $n=1$, this corresponds to
$M_1 \approx 8.3 \, \beta \Mpl$ and $M_2 \approx 0.09 \Mpl $,
both being sub-Planckian.
For $\xi = 0$, 
one has the borderline case  
$M_1=M_2= M_{\rm crit}$ 
with 
\begin{equation}
\Ma=\sqrt{2\beta} \, \Mpl=\sqrt{\alpha_e} \, n \, \Mpl \, .
\end{equation}
This corresponds to a 
lower bound for the size of the event horizon.
For $n=1$, one finds $M=M_\mathrm{crit} = 0.04 \Mpl$.

\begin{center}
\begin{figure}[b]
\includegraphics[scale=0.45]{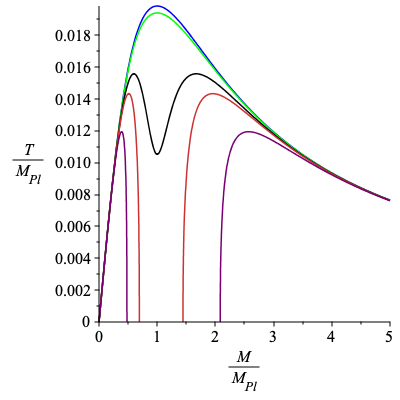}
\caption{ Hawking temperature curves for the GUP-RN black hole
with $\beta = 2$ and $n=10$ (blue), $n=16$ (green), $n=23$ (black), $n=25$ (red) and $n=30$ (purple).
The curves for $n>23$ have vanishing temperatures at the masses $M_1$ (left) and $M_2$ (right) defined by Eq.~\eqref{m1m2}. $T=0$ endpoints in the super-Planckian regimecorrespond to stable charged remnants.}
\label{PT_RNtemp}
\end{figure}
\end{center}

The Hawking temperature for the RN-GUP black hole is evaluated from the surface gravity as
\beq
kT 
= \frac{ \Mpl^2 \sqrt{\Ma^2 - \alpha_e n^2 \Mpl^2}}{2\pi  \, ( \Ma + \sqrt{\Ma^2 - \alpha_e n^2 \Mpl^2} \, )^2} \, .
\label{TKeq}
\eeq
Unlike the usual RN case, 
this specifies a temperature for both super-Planckian and sub-Planckian masses, whatever the value
of $n$. 
Figure \ref{PT_RNtemp} shows the $T(M)$ function for $\beta = 2$. A comparison  of the upper (lower) curves with Fig.~\ref{fig3} confirms the expected consequences of the BHUP correction.  However, as with the horizon curves, the extremal case introduces a discontinuity in the connection between
the sub-Planckian and super-Planckian regimes.
This shows that the  presence of extremal configurations has an important impact on the thermodynamics. 

We now discuss the different cases in more detail. The $\xi < 0$  case does not admit any extremal solutions and so one expects a  temperature profile very similar to that found in the neutral $M+1/M$ model,
the asymptotic behavior being
\begin{eqnarray}
T(M)\propto
\begin{cases}
M &  \mathrm{for} \ M\ll \Mpl\\
1/M & \mathrm{for}\ M\gg \Mpl \, .
\end{cases}
\end{eqnarray} 
 Indeed, Fig.~\ref{PT_RNtemp} reduces to Fig.~\ref{fig3}(a) in the $n=0$ limit, so we just recover the GUP-Schwarzschild temperature, as can also be seen from Eq.~\eqref{TKeq}.  
The $\xi > 0$ case has the same asymptotic limits but $T(M) \rightarrow 0$ for $M \to M_1^-$ and $M \to M_2^+$. 
The fact that the temperature also vanishes for $M\to 0$ and $M\to \infty$ 
implies the presence of two maxima for  the temperature, where phase transitions occur. 
For the $M>M_2$ branch, the phase transition is from a negative to a positive heat capacity regime, i.e. the prelude to the black hole SCRAM\footnote[1]{This
 is the cooling down phase during the final stages of evaporation, leading to a stable zero-temperature configuration. The term SCRAM, introduced in Ref.~\cite{nicolini09} and borrowed from nuclear reactor technology, is an acronym for ``Safety Control Rod Axe Man'', was coined by Enrico Fermi
during the Manhattan Project in 1942
and  still used to indicate the emergency shutdown of a
nuclear reactor.}. 
The $M < M_1$ branch  is anomalous because the extremal black hole with $M = M_1$ is no longer the end-point of the positive heat capacity (SCRAM) cooling phase. Rather it represents an unstable configuration with $T = 0$. Any perturbation of such a configuration (e.g. loss of charge via the Schwinger effect) would slightly increase the temperature before triggering a heating phase in a negative heat capacity regime. Such heating up can be dubbed anti-SCRAM and would terminate with a maximum temperature where a phase transition to positive heat capacity cooling takes place. The hole evaporates without leaving any remnant since $T$ and $M$ go to zero together.

For the $\xi = 0$ case, there is a remnant due to the double-zero of the temperature, 
\begin{eqnarray} 
T(M)\approx 0 &\quad& \mathrm{for}\ M\to M_\mathrm{crit}^\pm \, ,
\end{eqnarray} 
and this is expected to have both SCRAM and anti-SCRAM features. 
This means that 
a black hole with initial mass $M > \Mpl$  might not end up with mass $M_{\rm crit}$ but  follow the whole curve down to $M=0$.
The small and large oscillations with mass gaps  in Fig 7 
confirm this behaviour. We note that the double-zero case occurs only if $\beta$ is subject to a quantization rule: for $\xi = 0$, one obtains the value $\beta_{\rm min}$ given by Eq.~\eqref{mext3},
this relation establishing a  connection between the GUP and electrodynamics.
The temperature dip just reflects the fact that the black hole temperature goes to zero as one approaches the extremal solution.
The introduction of a Planck-scale oscillatory behaviour for increasing values of $n$
was previously observed  in the study of  
an extra-dimensional GUP-modified Schwarzschild spacetime and dubbed a {\it lighthouse effect} \cite{kkmn}.  
We conclude this section with some considerations of the Schwinger effect. The introduction of the mass parameter $\Madm$ modifies Eqs.~\eqref{eq:Coulomb} and \eqref{eq:Schwingertrigger}. 
With masses in Planck units, 
Eq.~\eqref{eq:Coulomb} becomes 
\begin{equation}
\frac{Q}{M+\beta/(2M)}<\frac{m}{q} \, .
\label{eq:CoulmodRN}
\end{equation}
This implies negligible Coulomb interaction for
\begin{equation}
M<\frac{qQ-\sqrt{q^2Q^2 -2\beta m^2}}{2m}\quad\mathrm{or}\quad M>\frac{qQ+\sqrt{q^2Q^2 -2\beta m^2}}{2m}
\end{equation}
for sub-Planckian and super-Planckian black holes, respectively. 
If $m$ is the mass of the electron, $m_e \sim 
10^{-22}$,
it is reasonable to assume $2\beta m^2\ll q^2 Q^2$, so that the discriminant in the square root is positive.  
Thus the inequality in the super-Planckian 
regime becomes equivalent to Eq.~\eqref{eq:Coulomb} with the same mass limits. On the other hand, the inequality in the sub-Plankian regime leads to the condition
\begin{equation}
M<\frac{\beta m}{2qQ} = \frac{\beta m}{2n \alpha_e} \, ,
\label{upper} 
\end{equation}
where we put $q=e$ at the last step. For $n\sim \beta \sim 1$ and $m = m_{e}$, the Coulomb interaction therefore dominates over gravity for masses in the range from $10^{-28}$ kg to $10^{12}$ kg. This can be compared with the mass range implied by Eq.~\eqref{eq:Schwingertrigger},
\begin{equation}
\frac{Q}{r_+^2}>\frac{Q}{4[ M+\beta/(2M)]^2}\geq \frac{\pi m^2}{e} \, .
\label{eq:SchwingermodRN}
\end{equation}
In the super-Planckian regime, one obtains the same condition as from Eq.~\eqref{eq:Schwingertrigger}. In the sub-Planckian
regime, the inequality implies
\begin{equation}
M\geq\frac{\beta}{2}\sqrt{\frac{\pi}{eQ}}\ m  = \frac{\beta}{2}\sqrt{\frac{\pi}{n \alpha_e}}\ m \, .
\label{lower} 
\end{equation}
For $n\sim \beta \sim 1$ and $m = m_{e}$, this
corresponds to $M \gtrsim 10^{-29}$ kg.
This bound is
slightly larger than  the mass of the produced particle (i.e. the electron)
because the size of 
a sub-Planckian black hole
is its Compton wavelength.
However,  the extremal case $M_{\rm ADM} = Q$ leads to even stricter limits. In the super-Planckian regime, one finds the result of the previous section, $M < 10^{35}$~kg, but in the sub-Planckian regime, the bound becomes
\begin{equation}
M > \frac{ \beta \pi m^2}{2 e} \, .
\end{equation}
For $\beta \sim 1$ and $m = m_e$, one has $M > 10^{-43} \Mpl  \sim 10^{-51}$~kg, a limit that reveals the
 implausibility  of extremal stable configurations. 
We conclude that  sub-Planckian black holes have relevant electrodynamics effects for masses exceeding that of the electron. They undergo a sudden discharge via both standard evaporation and Schwinger emission.
Only for masses below $10^{-28}$ kg can they 
retain electric charge. 

From this viewpoint, charged black holes, described by either the 
RN or GUP-RN solutions,
 are transient states. They may be
produced in the early Universe but they will
decay to neutral configurations quite rapidly. In GUP case,
 the metric \eqref{newmetric} represents the ground state in the black hole parameter space.
 In contrast to the RN
case, however, both Eqs.~\eqref{upper} and \eqref{lower} allow
 the identification of elementary particles with
black holes. 
A hypothetical sub-Planckian black hole with $M = m_e$ would not undergo rapid discharge since 
both Coulomb and Schwinger effects are negligible for such a mass.
Therefore, apart from spin effects, the electron might be interpreted as a sub-Planckian black hole in its non-extremal configuration.

\section{Self-Completeness and GUP-modification of Kerr black holes}

Following our analysis of the RN and GUP-RN black hole solutions, we now turn to the Kerr solution, 
first reviewing its standard classical features.
The metric for a Kerr black hole  of mass $M$ and angular momentum $J$ is
\beq
ds^2 = \left(1-\frac{r_{\rm S} r}{\rho^2}\right) dt^2 - \frac{\rho^2}{\Delta} dr^2 - \rho^2~ d\theta^2 - \left(r^2 + a^2  + \frac{r_{\rm S} r a^2}{\rho^2} \sin^2\theta\right) \sin^2\theta~ d\phi^2+ 
\frac{2 r_{\rm S} r a \sin^2\theta}{\rho^2} ~dt~d\phi \, ,
\eeq
where $r$ is the spheroidal radial coordinate	 and
\beq
r_{\rm S} = 2M/\Mpl^2~~,~~a = \frac{J}{M}~~,~~\rho^2 = r^2 + a^2 \cos^2\theta~~,~~\Delta = r^2 - r_{\rm S} r + a^2 \, .
\eeq
The horizon structure is more complicated than in the RN case, since the spin introduces a non-spherical ergosphere region.  For present
purposes, however, we will restrict attention to the outer and inner horizons, defined by
\beq
\Delta = 0~~~\Longrightarrow~~~r_{\pm}  = \frac{M}{\Mpl^2} \left(1\pm \sqrt{1-\frac{a^2\Mpl^4}{M^2}}\right) \, .
\label{koh}
\eeq
For a black hole which is far from extremal,  this gives
\bea
r_{\pm} \approx 
\begin{cases}
\frac{2M}{\Mpl^{2}} \left(1 - \frac{\gamma' \Mpl^4}{M^4}\right) & (+) \\ 
\frac{2\gamma' \Mpl^2}{M^3}\left(1 + \frac{\gamma' \Mpl^4}{M^4}\right) & (-)
\end{cases} \, ,
\eea
where  $\gamma' \equiv n^2/4$ (different from   $\gamma$ in the RN case by a factor of $\alpha_e$) and $J = n$ (in units with $\hbar =1$). 
The first expression is no longer of the form given by Eq.~\eqref{unified} but corresponds to the quadratic version of the GUP \cite{CMP}, while the second expression differs from the Compton wavelength. 

The extremal case corresponds to the spectrum of masses,
\beq
M = \sqrt{n} \, \Mpl \, .
\label{ek}
\eeq
On the other hand, the condition $r_{\rm C} = r_+$  (required for self-completeness) implies
\bea
\frac{1}{M} = \frac{M}{\Mpl^2} \left(1\pm \sqrt{1-\frac{a^2\Mpl^4}{M^2}}\right) \Rightarrow 
M  =  
 \Mpl\sqrt{\frac{1+n^2}{2}} \, .
\label{sck}
\eea
So the analysis is  similar to the RN case and the qualitative features of Fig.~\ref{fig12} still apply. 
However, the Compton line intersects the outer horizon for all values of $n$ (eg. at $\Mpl /\sqrt{2}$ for $n=0$ and $\Mpl$ for $n=1$).  
This contrasts with the RN case, where $n$ could not exceed  $1/\sqrt{\alpha_e} \approx 11$ for the intersect with $r_+$ and $\sqrt{2 /\alpha_e} \approx 16$ for the intersect with $r_-$.
The temperature can be shown to be
\beq
T = \frac{1}{4\pi} \frac{r_+ - r_-}{r_+^2+(n/M)^2} \, ,
\eeq
which vanishes for the extremal solutions given by Eq.~\eqref{ek}. 

The evaporation of the black hole determines a spin-down process. Due to the conservation of the angular momentum, the emitted particle has to have a spin aligned with the angular momentum of the black hole. The spin-down also occurs 
because of the superradiant modes scattered by the hole, an effect known as Starobinsky-Unruh radiation \cite{Starobinsky:1973aij,Unruh:1974bw}. In much the same way as for the charged case in the previous section, metric \eqref{newmetric} turns out to be the ground state.
However, the decay of rotation is
expected to be slower than 
the discharge \cite{carter}.
These considerations have interesting implications for the link between elementary particles and black holes.  
While black holes can match elementary particles at the end of the spin down phase, the spin of those formed
by particle collisions 
can be at the most the sum of the spin of the particles. For example, 
for black holes created by the collisions of spin $1/2$ particles, the total angular momentum will be $\pm 1$ or $0$, depending on the spin alignment.
Particles with higher spin can be conjectured (eg. the Rarita-Schwinger field \cite{Rarita:1941mf} of the spin-3/2 gravitino) but these would form a black hole only if anti-aligned and this also applies for higher spin fields in string theory. 

\begin{center}
\begin{figure}[h!]
\includegraphics[scale=0.5]{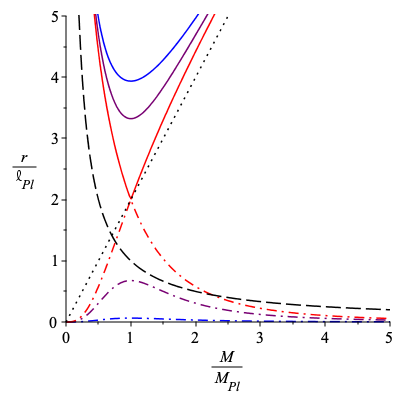}
\caption{The outer (solid) and inner (dash-dot) horizons for the GUP-Kerr black hole with $\beta =2$ and $n=1$ (blue), 3 (purple) and 4 (red).  The $n=4$ curves have a discontinuity at $M=\Mpl$ and $r=2 \ellp$, corresponding to a phase transition.
The dashed line is the Schwarzschild radius 
and the dotted curve is the
Compton wavelength.} 
\label{KerrH1}
\end{figure}
\end{center}

\begin{center}
\begin{figure}[h!]
\includegraphics[scale=0.5]{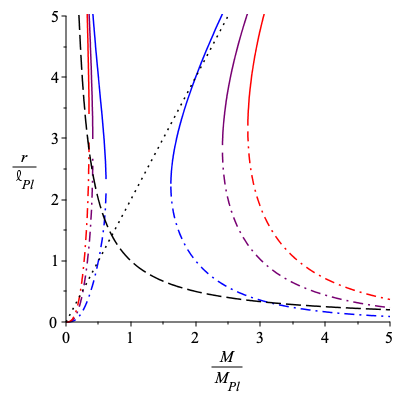}
\caption{The outer (solid) and inner (dash-dot) horizons for the GUP-Kerr black hole with $\beta =2$ and $n=5$ (blue), 8 (purple) and 10 (red), showing the mass gaps for $n>4$. The dotted line is the Schwarzschild radius, $r=2M/\Mpl^2$, and the dashed curve is the Compton wavelength.} 
\label{KerrH2}
\end{figure}
\end{center}

\begin{center}
\begin{figure}[h!]
\includegraphics[scale=0.5]{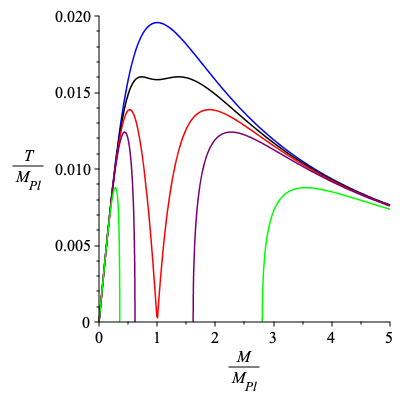}
\caption{Hawking temperature for the GUP-Kerr black hole 
 for $\beta=2$.  Curves correspond to the parameters $n=1$ (blue), 3 (black), 4 (red), 5 (purple) and 10 (green).  The critical value $n=4$ introduces a phase transition, similar to that for
the extremal RN-GUP black hole, beyond which there is a mass gap with no black holes. $T=0$ endpoints in the super-Planckian regime correspond to stable spinning remnants.}
\label{KerrT}
\end{figure}
\end{center}

The Kerr metric can be modified to include the GUP-modified mass $\Ma$
term by changing 
$\Delta$ to
\beq
\Delta = r^2 - \frac{2 \Ma r}{\Mpl^2}+\left(\frac{n}{\Ma}\right)^2 \, .
\eeq
The condition $\Delta = 0$ then gives outer and inner horizons at
\beq
r_\pm = \frac{\Ma}{\Mpl^2} \pm \sqrt{\frac{\Ma^2}{\Mpl^4}-\frac{n^2}{\Ma^2}}~~
\label{KGUP}
\eeq
and their form as functions of $M$ is shown in Figs.~\ref{KerrH1}
and \ref{KerrH2} for $\beta =2$.  For $n < 2\beta$, the horizons have a similar form to that shown in Fig.~\ref{rnhorizon}. Since $r_+$ is a monotonic function of $\Ma$, it has a minimum at the minimum at $\Ma$, corresponding to 
\beq
M = M_{\rm crit} = \sqrt{\beta/2} \, \Mpl  \, , \quad  r_+ = [\sqrt{2 \beta} + \sqrt{ 2 \beta - n^2/(2 \beta)} \, ]  \,  \ellp \, .
\eeq
Similarly $r_-$ has a maximum at 
\beq
M = M_{\rm crit} = \sqrt{\beta/2} \, \Mpl  \, , \quad  r_- = [\sqrt{2 \beta} - \sqrt{ 2 \beta - n^2/( 2\beta} \, ]  \,  \ellp \, .
\label{mcrit}
\eeq
These features are  illustrated in Fig.~\ref{KerrH1}. 
The minimum and maximum merge for the extremal solution ($r_+ = r_-$) when $n = 2\beta$. For $n > 2\beta$, the minimum and maximum no longer exist and the horizons have the form shown in Fig.~\ref{KerrH2}, which might be compared to Fig.~\ref{PT_RNH}. There are horizons for $M>M_{+}$ and $M < M_{-}$ where
\beq
M_\pm = \frac{\Mpl}{2} \left(\sqrt{n} \pm\sqrt{n-2\beta}\right) \, ,
\label{kerrex}
\eeq
these values corresponding to the extremal condition $\Ma = \sqrt{n} \, \Mpl$. Thus solutions
with $n > 2\beta$ exists in two self-complete phases, representing 
 either sub-Planckian or super-Planckian 
black holes. The transition at $n = 2\beta$ resembles  that for the RN-GUP black hole shown in Fig.~\ref{rnhorizon}, with the black hole bifurcating into two separate solutions with dual masses $M$ and $\Mpl^2/M$.  
The temperature is 
\beq
T = \frac{1}{4\pi} \frac{r_+ - r_-}{r_+^2+(n/\Ma)^2}
\eeq
and its behaviour as a function of $M$ is indicated in Fig.~\ref{KerrT}. It vanishes for the values of
  $M$ given by Eqs.~\eqref{mcrit} and \eqref{kerrex}.


\section{Conclusions}

 In this paper, we have studied the self-complete behaviour and thermodynamics of both the Reissner-Nordstr\"om (RN) and Kerr black holes, as well as their GUP-modified versions.
For each solution, we have shown that self-completeness ({\it i.e.} the condition that the Compton line intersects the outer black hole horizon) introduces a spectrum of 
minimum masses providing the charge or spin 
parameter is not too large. 
For self-completeness the  maximum value of $n$ is $11$ for RN but there is no limit on $n$ for Kerr.

For the GUP-modified  versions of the RN metric, we have shown that the outer horizon behaves as in 
 the GUP-Schwarzschild case \cite{CMN}, 
thus providing a continuous transition between the gravitational ($r_{\rm CS} \propto M$) and Compton ($r_{\rm CS}\propto M^{-1}$) scaling.  As $n$ increases, the smoothness of the 
 minimum becomes sharper until the charge 
 reaches a maximum value of $n$ which depends on $\beta$ but differs from the value of $11$ for the self-complete RN case.
Beyond this critical value,
 there is a
 phase transition that introduces a mass gap.  This replicates the behaviour of the standard RN inner and outer horizons, so we have speculated that these solutions represent a super-Planckian black hole on the right and a sub-Planckian black hole or 
Compton-like object ({\it i.e.} a particle) on the left.

We have also demonstrated that the GUP-RN black hole acquires a temperature 
similar to the GUP-Schwarzschild one, with a zero-temperature remnant for some range of charge up to the maximum value of $n$.
The temperature also exhibits Planck-scale oscillatory behaviour for decreasing $M$, similar to the well-known `lighthouse effect'~\cite{kkmn}.   
This phase transition 
 is also evident in the $M$-dependence of the Hawking temperature.
The GUP-Kerr metric exhibits similar behaviour
but there is a critical spin  
instead of a critical charge
  and this determines the subsequent  phase transition.

In both the charged and rotating cases, 
the black hole characteristics in the sub-Planckian regime correspond
 to dimensional reduction 
from $(3+1)$-D to an effective $(1+1)$-D spacetime, as with the GUP-Schwarzschild solution \cite{CMN}.
Indeed, an interesting feature of the conventional 
RN and Kerr metrics,
 previously unnoticed as far as we are aware,
is that the inner
 horizon radius for far-from-extremal RN black holes scales as $1/M$, which is 
the same relation for the horizon of a $(1+1)$-D black hole \cite{jmplb12}.
This also applies for the outer horizon in the sub-Planckian part of the BHUP solution.
 Dimensional reduction is an expected feature of a final theory of quantum gravity \cite{thooft}, so we suggest that this feature indicates that gravity within the horizon may itself be lower-dimensional.

We should also mention some  general conceptual implications of this work. One prediction of this paper is that there is a fundamental link between elementary particles and black holes. This proposal goes back to the 1970s, when it was motivated in the context of strong gravity theories by the link between Regge trajectories and extreme Kerr solutions. In our case,  it is prompted by the
Compton-Schwarzschild correspondence, which is based on the $M \rightarrow 1/M$ duality. Indeed, this suggests that elementary particles could be black holes with sub-Planckian mass. However, this duality no longer applies in the charged and rotating cases since the Compton wavelength is independent of $Q$ and $J$ for a particle. 
So do we just drop this duality or do we modify the electrostatic term in the RN solution and the angular momentum term in the Kerr solution in such a
way that it is preserved? We have argued against this and the presence of mass-gap solutions  confirms that large charge and/or
 angular momentum destroys this duality but perhaps this issue requires further consideration. 

If there is a link between  elementary particles and black holes, what is the evidence for this and what are the implications? One puzzling feature is that no fundamental particles have charge exceeding $e$ or spin exceeding $2 \hbar$.  By contrast  self-completeness implies that standard RN  has a maximum charge of 11 or even 16 if one includes solutions in which the inner horizon intersects the Compton line. Although super-Planckian black holes necessarily discharge through the Schwinger mechanism, we have shown that sub-Planckian ones in the GUP-modified RN solution can maintain their charge for sufficiently small masses.  

 There is also the important issue of  the interpretation of the mass $M$
 in gravitational theory and how it should be defined in the sub-Planckian and super-Planckian regimes.  Its usual role is to instruct spacetime how to curve but GUP effects change the instructions, so that the standard Schwarzschild, RN and Kerr metrics no longer apply. For large classical black holes, $M$ is the ADM mass, while for particles it is related to the centre-of-mass energy, as calculated for particle collisions. 
In between these limits, all we know is that 
 the physical mass around
the Planck scale has to be some mixture of these two definitions.
This inevitably reflects the fact that the Planck scale is the confluence of the scales of gravity and quantum mechanics. While the exact expression is uncertain,
we have argued that
Eq.~\eqref{unified} is at least a consistent
amalgamation.

Finally, this work is important for the concept of  ``gravitational self-completeness''. This concept has a long history and has the important implication that 
any attempt to probe a particle above the Planck energy will result in the formation of a black hole. Most minimal-length mechanisms include this feature. However, if there is a duality between particles and black holes, this also has the implication that any attempt to produce a black hole below the Planck length will probe the Compton scale instead, so that the singularity at the centre of a black hole is inaccessible. 
But is self-completeness the notion that experiments cannot go below the Planck scale or does it also imply a distinction between particles and black holes, corresponding to some critical point in the ($M,R$) diagram)? In the latter case, our `$M+1/M$' model solutions would not qualify. However,  they still have the feature that one cannot probe below the Planck length, so we would advocate extending the definition of self-completeness to include this case.

\vskip 0.5cm

\noindent{\bf Acknowledgments}\\ JM and PN would like to thank the generous hospitality of Queen Mary University of London, at which this work was done.  The work of JM and HM was supported in part by a Frank R. Seaver Summer Research Fellowship from Loyola Marymount University. 
The work of PN has been partially supported by GNFM, the Italian National Group for Mathematical Physics.  We thank R.~Balbinot, G.~Gibbons and E.~Spallucci for useful comments.


\end{document}